\documentclass{emulateapj}
\usepackage{epsfig}
\usepackage{multirow}
\usepackage{graphicx}
\usepackage{epstopdf}
\usepackage{amsmath, amsthm, amssymb}
\usepackage{natbib}
\DeclareGraphicsExtensions{.pdf,.eps,.eps.gz}
\tightenlines

\newcommand \lsim{\mathrel{\rlap{\lower4pt\hbox{\hskip1pt$\sim$}}
    \raise1pt\hbox{$<$}}}
\newcommand \gsim{\mathrel{\rlap{\lower4pt\hbox{\hskip1pt$\sim$}}
    \raise1pt\hbox{$>$}}}

\newcommand     \Angstrom       {\,{\rm \AA}}

\newcommand     \cm     {\,{\rm cm}}

\newcommand     \kms    {\,{\rm km~s}^{-1}}
\newcommand     \kpc    {\,{\rm kpc}}

\newcommand     \pc     {\,{\rm pc}}

\newcommand{\beq}{\begin{equation}}
\newcommand{\eeq}{\end{equation}}
\newcommand{\beqa}{\begin{eqnarray}}
\newcommand{\eeqa}{\end{eqnarray}}

\newcommand{\htwo}      {H$_2$}

\def\HI{\ion{H}{1}}
\def\HII{\ion{H}{2}}
\def\NI{\ion{N}{1}}
\def\NaI{\ion{Na}{1}}
\def\AlII{\ion{Al}{2}}

\def\OI{\ion{O}{1}}

\def\SII{\ion{S}{2}}
\def\SIII{\ion{S}{3}}
\def\SiII{\ion{Si}{2}}

\def\FeII{\ion{Fe}{2}}
\def\FeIII{\ion{Fe}{3}}
\def\ArI{\ion{Ar}{1}}

\newlength{\figwidth}
\countdef\decade=200
\advance\decade by \year
\countdef\hours=201
\advance\hours by \time
\divide\hours by 60
\countdef\mins=202
\advance\mins by \hours
\multiply\mins by 60
\multiply\hours by 100
\countdef\miltime=203
\advance\miltime by \hours
\advance\miltime by \time
\advance\miltime by -\mins

\begin{document}

\title{A Low Metallicity Molecular Cloud In The Lower Galactic Halo\footnotemark[1]}

\author{Audra K. Hernandez}
\affil{Department of Astronomy, University of Wisconsin, 475 North Charter Street, Madison,  WI 53706, USA;\\ hernande@astro.wisc.edu}
\author{Bart P. Wakker}
\affil{Department of Astronomy, University of Wisconsin, 475 North Charter Street, Madison,  WI 53706, USA;\\ wakker@astro.wisc.edu}
\author{Robert A. Benjamin}
\affil{Department of Physics, University of Wisconsin-Whitewater, Whitewater, WI, USA;\\
 benjamin@astro.wisc.edu}
\author{David French}
\affil{Department of Astronomy, University of Wisconsin, 475 North Charter Street, Madison,  WI 53706, USA;\\ 
frenchd@astro.wisc.edu}
\author{Juergen Kerp}
\affil{Argelander-Institut f\"{u}r Astronomie, Auf dem H\"{u}gel 71, 53121 Bonn, Germany;\\ 
jkerp@astro.uni-bonn.de}
\author{Felix J. Lockman}
\affil{National Radio Astronomy Observatory, Green Bank, WV 24944, USA;\\ 
fjlockman@nrao.edu}
\author{Simon O'Toole}
\affil{Australian Astronomical Observatory, PO Box 296, Epping 1710, Australia;\\ 
otoole@aao.gov.au}
\author{Benjamin Winkel}
\affil{Max-Planck-Institut f\"{u}r Radioastronomie (MPIfR), Auf dem H\"{u}gel 69, 53121, Bonn, Germany;\\ 
bwinkel@mpifr.de}

\begin{abstract}

We find evidence for the impact of infalling, low-metallicity gas on the Galactic disk.  This is based on FUV absorption line spectra, 21-cm emission line spectra, and FIR mapping to estimate the abundance and physical properties of IV21 (IVC135+54-45), a galactic intermediate-velocity molecular cloud (IVMC) that lies $\sim300$ pc above the disk. The metallicity of IV21 was estimated using observations toward the sdB star PG1144+615, located at a projected distance of 16 pc from the cloud's densest core, by measuring ion and \HI~ column densities for comparison with known solar abundances. Despite the cloud's bright FIR emission and large column densities of molecular gas as traced by CO, we find that it has a sub-solar metallicity of $\rm \log (Z/Z_{\odot})=-0.43\pm0.12$~dex.  IV21 is thus the first known sub-solar metallicity cloud in the solar neighborhood.  In contrast, most intermediate-velocity clouds (IVC) have near-solar metallicities and are believed to originate in the Galactic Fountain.  The cloud's low metallicity is also atypical for Galactic molecular clouds, especially in the light of the bright FIR emission which suggest a substantial dust content. The measured I$_{100\mu m}$/N(\HI) ratio is a factor of three below the average found in high latitude \HI~ clouds within the solar neighborhood.  We argue that IV21 represents the impact of an infalling, low-metallicity high-velocity cloud (HVC) that is mixing with disk gas in the lower Galactic halo.

\end{abstract}


\section{Introduction} 
Identified as dense concentrations of neutral hydrogen (\HI) within the galactic halo, high- and intermediate-velocity clouds (HVCs and IVCs) have radial velocities ($\rm |v_{LSR}| \ge 100~\kms$ and $\rm |v_{LSR}| \sim 30-100~\kms$, respectively) that are incompatible with differential Galactic rotation \citep{Wakker1991}.  Earlier studies found that HVCs have varying chemical compositions \citep{Wakker1999, Gibson2001, Tripp2003, Sembach2003, Fox2005}, indicating that the clouds may have different origins. Generally, it is thought that these objects are infalling, either as vertical galactic circulation (i.e. the Galactic Fountain; \citealt{ShapiroField1976,Bregman1980}) or as clouds accreting from intergalactic space \citep{Oort1970}.  Regardless of their origins, HVCs and IVCs play an integral role in the evolution of our Milky Way Galaxy.  

\footnotetext[1]{Based on observations made with the NASA/ESA Hubble Space Telescope, obtained from MAST at the Space Telescope Science Institute, which is operated by the Association of Universities for Research in Astronomy, Inc., under NASA contract NAS 5-26555. These observations are associated with program No. 12275.  The Green Bank Telescope is part of the National Radio Astronomy Observatory is a Facility of the National Science Foundation, operated by Associated Universities, Inc. }

\citet{BenjaminDanly1997} suggested that an infalling cloud will decelerate due to gas drag as it approaches the Galactic disk.  Consequently, intermediate velocity clouds would lie at intermediate distances above the plane of the galaxy, while high velocity clouds would lie further out.   If true, studies of IVCs can provide key insight into the interstellar environment a few hundred parsecs above the disk, otherwise referred to as the lower Galactic halo. As a cloud decelerates toward the Galactic plane, dynamical information regarding its origin will be lost.  Previous work on peculiar velocity gas at high latitudes shows that many HVCs have sub-solar metallicities $\rm (\log (Z/Z_{\odot})=0.1-0.2~dex)$, are located 5-10 kpc above the disk \citep{Wakker1999,Fox2004,Wakker2007}, and likely are due to the infall of extragalactic gas. On the other hand, IVCs tend to be at $\rm z=0.5-1 ~\kpc$ \citep{Wakker2001}, have near-solar metallicity \citep[e.g.][]{Richter2001a,Richter2001b,Sembach2004}, and are interpreted as evidence for a Galactic Fountain.  Thus, the best method to determine the origin of an HVCs/IVCs is to measure its metallicity.

At the northern disk-halo interface lies the intermediate-velocity molecular cloud (IVMC) named IV21 (IVC135+54-45) centered at $l=135.3^{\circ}$, $b=52^{\circ}$ \citep[][]{KuntzDanly1996,Weiss1999}.  IV21 does not appear to be part of the solar metallicity IV-Arch, which is at a distance of $\rm z=0.8-1.5~\kpc$ \citep{Wakker2001}, even though it is located in the same region of the sky and has similar LSR velocities.  \citet{Benjamin1996} estimated the distance of IV21 to be $d=355\pm95$~pc based on intermediate-velocity \NaI~ absorption detections towards the nearby stars HD101714 and HD101075 and a lack of a \NaI~ detection  toward the neighboring star BD $+63^{\circ}985$.  The distances to all these stars were estimated using the Schmidt-Kaler spectral type absolute-magnitude relationship, assuming no correction for extinction.  This distance combined with its high galactic latitude ($b=54^{\circ}$) places IV21 at a height of $z=285\pm75$~pc \citep{Benjamin1996}.   IV21 is one of only eight known IVMCs, a subset of IVCs known to contain high column densities ($\rm \sim 10^{20} ~cm^{-2}$) as indicated by CO emission in their densest regions \citep{Weiss1999, MagnaniSmith2010}.  Based on studies of the cloud's emission, much is already known about its physical characteristics.  Figure \ref{grid1} presents the IRAS 100 ~$\mu$m map indicating three concentrations of infrared cirrus clouds.  CO detections within the IVMC's densest regions were first made by \citet{Heiles1988}. By combining both the IRAS and \HI~ emission line data, \citet{Weiss1999} found a linear relationship between the 100 ~$\mu$m emission and the \HI~ column density of I$_{100 \mu m}$/N(\HI)$= 0.98\times10^{-20} \rm ~(MJy~sr^{-1}~\cm^{2})$.  This correlation is in agreement with the results of \citet{BoulangerPerault1988} who estimated a ratio of I$_{100\mu m}$/N(\HI)$= 0.92\times10^{-20} \rm ~(MJy~sr^{-1}~\cm^{2})$ for high latitude clouds (b$>50^{\circ}$) within the solar neighborhood.  \citet{Weiss1999} also found that IV21 contains two regions in different evolutionary states in the transition from atomic to molecular gas.  There is also evidence that this cloud has an X-ray shadow against emission from the Galactic halo \citep{Snowden1994, Benjamin1996, Kappes2003}, confirming that it is relatively local. 

In this paper we estimate the metallicity of IV21 using absorption line spectra taken with STIS and FUSE toward the subdwarf B (sdB) star PG1144+615 (also referred to as Feige 48).  This background point source lies at a heliocentric distance of $\sim1~\kpc$ with a projected distance of 16.1 pc from the center of the cloud's densest core (Fig. \ref{grid1}).  Originally classified as a ``faint blue star'' by \citet{Feige1958}, PG1144+615 is believed to have a total mass similar to the canonical sdB value of 0.46 M$_{\sun}$ \citep{Charpinet2005}.  Through a combination of high S/N ratio Keck HiRES spectra and line-blanketed LTE and NLTE atmospheric models, \citet{Heber2000} determined the physical parameters of the star, finding $T_{eff}=29500\pm300$ K, $\log g=5.50\pm0.05$ and $\rm \log N(\rm He)/N(\rm H) = -2.93\pm0.05$.    Later, \citet{Otoole2004} confirmed the projected rotational velocity upper limit of $V \sin i \leq 5~\kms$ using a combination of \FeIII~ absorption line spectra from both HST/STIS and archival FUSE spectra. This study is also responsible for discovering the binary nature of PG1144+615 with a period of $P=0.376\pm0.003$~d and a semi-amplitude velocity of $\rm K=28.0\pm0.2 ~\kms$.  Furthermore, \citeauthor{Otoole2004} suggested that the companion star is most likely a white dwarf given the system's nearly pole-on inclination and small velocity semi-amplitude.  

Using 23 FUV interstellar absorption lines in the direction of PG1144+615 we estimated the column densities of eight ions within IV21, including \OI, \SII, \SiII, and \FeII.  Furthermore, we determined the IVMC's chemical abundances by estimating the atomic hydrogen column density, N(\HI), in the direction of PG1144+615 using 21-cm line observations taken with the 100-m Robert C. Byrd Green Bank Telescope \citep[GBT,][]{Prestage2009}.  Additionally, we used IRAS 100 $\mu$m emission maps to assess the small scale structure within the GBT beam to achieve a more accurate estimate of N(\HI). We compared these estimated chemical abundances to the known solar abundances in \citet{Asplund2009}, finding that IV21 has a $\rm \log (Z/Z_{\odot})=-0.43\pm0.12$~dex metallicity, making it the \textit{only} known cloud in the solar neighborhood known to have a significantly sub-solar metallicity.  Section 2 presents the observational data sets used, while Section 3 discusses our methods used to derive the ion, \HI~, and $\rm H_2$ column densities, and ion abundances.  In Section 4 we discuss the possible causes and implications for the presence of such a metal-poor cloud within the solar neighborhood. 

\section{Observational Data Sets}
\subsection{FUV Data}
Absorption spectroscopy was taken for the sdB star PG1144+615 using both the Far Ultraviolet Spectroscopic Explorer (FUSE) (3 April 2002, program C036, PI Benjamin) and the Space Telescope Imaging Spectrograph (STIS) (13 May 2001, program 8635, PI Heber).   FUSE observations consist of eight segments, covering different, but overlapping, wavelength ranges.  We used the segments LiF1A (987-1082 $\Angstrom$), LiF1B (1094-1187 $\Angstrom$), LiF2A (1087-1181 $\Angstrom$), LiF2B (979-1075 $\Angstrom$), and SiC2A (917-1006 $\Angstrom$).  Individual observations were taken in time-tag mode with the LWRS aperture and an exposure time of $\sim$20,000 sec.  Calibrations were done using CalFUSE (v.2.4.0).  Individual observations were stacked prior to being analyzed using the wavelength adjustment routine described in \citet{Wakker2003}. The final stacked FUSE spectra had a spectral resolution of $20\kms$, or 64.1 m$\Angstrom$.  STIS observations were taken in time-tag mode with an exposure time of 3500 seconds. The echelle gratings provide a wavelength coverage of $1171-1709 \Angstrom$ (E140M) and $1607-2366 \Angstrom$ (E230M).  Each STIS observation was retrieved from the STScI MAST server.  Prior to retrieval, each observation was reduced using CalSTIS v2.2.3.  The final stacked STIS spectrum has a spectral resolution of $6.5\kms$, or 30.6 m$\Angstrom$. 
 
\subsection{21-$\cm$ Data}
To obtain N(\HI) measurements in the direction of PG1144+615, we obtained a $0.5^{\circ}\times0.5^{\circ}$ grid of 21-cm spectra taken with the 100-m Robert C. Byrd Green Bank Telescope \citep[GBT,][]{Prestage2009}.  At 21-cm, the GBT has an angular resolution of 9\arcmin.1.  Data was taken using in-band frequency switching leading to a velocity coverage of $\pm500~\kms$ and a velocity resolution of $0.16~\kms$.  All spectra were calibrated and corrected for stray radiation using the methods described in \citet{Boothroyd2011}. The data were filtered with a hanning smoothing function for a final velocity resolution of 0.64~$\kms$ and spatial resolution of $1.75\arcmin$. The mean rms noise is 0.05 K in each $0.64~\kms$ channel. The GBT has typical system temperatures of 18 K at zenith. 

To study the large scale \HI~ structure of IV21 we used new 21-$\cm$ line observations of the entire cloud region taken with the Effelsberg 100-m telescope as part of the Effelsberg Bonn \HI~ Survey (EBHIS, \citealt{Winkel2010}, \citealt{Kerp2011}).  Figure \ref{grid1} presents a \HI~ column density map covering a $5^{\circ}\times5^{\circ}$ region.  The EBHIS is an northern all-sky survey (north of $\rm Dec=-5^{\circ}$) of the Milky Way and the extragalactic environment out to $\rm z\sim0.07$.  At 1.4 GHz,  the beam size is $9\arcmin.4$ and the velocity resolution is 1.49 km/s with an rms noise of $\sim90$ mK.   After gridding, the effective beam size is $10\arcmin.7$ and the velocity channel width is 1.29~$\kms$.  

We also used \HI~ maps provided by the Leiden/Argentine/Bonn Galactic \HI~ Survey \citep[LAB,][]{Kalberla2005} to view the extended environment of IV21.  The LAB survey provide full-sky \HI~ emission line observations with a spatial resolution of 36\arcmin~ and a velocity resolution of $1.3 ~\kms$.

\subsection{Infrared}
We use a 100~$\mu$m far-infrared (FIR) image taken by the Infrared Astronomical Satellite (IRAS).  The $5^{\circ}\times5^{\circ}$ map (Fig. \ref{grid1}a)  was downloaded from the IRAS Sky Server Atlas (ISSA).  At 100~$\mu$m IRAS provides an angular resolution of $3.0\arcmin\times5.0\arcmin$, a pixel scale of 1.5\arcmin, and a noise level of $\rm 0.07 ~(MJy ~sr^{-1})$.  To search for star formation and even finer cloud structure we obtained a 22~$\mu$m mid-infrared (MIR) image taken by the Wide-field Infrared Survey Explorer (WISE) at an angular resolution of 12\arcsec ~at 22~$\mu$m and a pixel scale of 1.5\arcsec ~(Fig. \ref{grid1}b).

\section{Determination of IV21 Column Densities and \htwo~Parameters}
\subsection{Column Densities of Metal Lines} 
We identified 23 interstellar absorption lines for eight ions, including \SII, \SiII, \FeII, and \OI.  The ion column densities were first estimated using the apparent optical depth (AOD) method from \citet{SavageSembach1991} and \citet{SembachSavage1992}.  However, we found that this method was inappropriate for these sightlines due to unresolved saturation.  When few lines are available for individual ions this method is more convenient for deriving column densities than other methods such as a curve-of-growth analysis or detailed profile fitting routines.  The S/N ratio of $\sim30$ for FUV spectra for PG1144+615 is more than adequate for use with the AOD method which was designed for high signal-to-noise spectra ($\rm S/N \ge 20$). In spectra with low S/N, on the other hand, there are systematic errors due to the relationship between apparent optical depth and the observed flux leading to an overestimated column density \citep{Fox2005}.
  
In our second trial, to avoid the significant errors of the AOD method, we used a curve-of-growth analysis to better estimate the final true column density of each ion.  Using a $\chi^2$ minimization test (Eqn. \ref{chi}), a best-fitting FWHM and column density were estimated for each ion with three or more resolved absorption lines using their measured equivalent widths. 
\begin{equation}
\chi^2=\Sigma \frac{[EW_{\rm obs}-EW_{\rm pred}(N,FWHM)]^2}{\sigma(EW)^2}.
\label{chi}
\end{equation}
The best-fitting column density and the FWHM for each atomic species and each \htwo~ J-level were then found by minimizing $\chi^2$. The measured equivalent widths for each absorption line are presented in Table \ref{ews}.  A final FWHM estimate was then calculated by taking a uniform weighted average of the best fitting FWHM found for \FeII~ and \SII.  
Subsequently a $\chi^2$ minimization test was done for each ion using the mean FWHM and measured equivalent widths to estimate a final column density, N$_{\rm cog}$.  

\begin{table}[!hb] 
\scriptsize
\caption{Interstellar Absorption Line Equivalent Widths}
\begin{center}
\begin{tabular}{lcc}
\tableline\tableline
Ion & $\lambda$ $(\rm \AA)$ & EW ($\kms$)\\
(1) & (2) & (3) \\
\tableline 
NI    & 1199.5 & $109.7\pm10.6  $\\
NI    & 1200.2 & $90.4\pm7.0       $\\
NI    & 1200.7 & $115.4\pm9.5    $\\
NI    & 1134.9 & $99.1\pm11.0      $\\
NI    & 1134.1 & $59.3\pm6.5       $\\
\tableline
OI    & 1039.2 & $95.7\pm13.4$ \\
OI    &  976.4  & $41.3\pm10.5$  \\
OI    &  936.6  & $63.7\pm17.6$  \\
\tableline
SiII  & 1304.3 & $91.6\pm13.5$  \\
SiII  & 1020.6 & $49.5\pm7.0$  \\
\tableline
AlII  & 1670.7 & $116.4\pm14.1$ \\
\tableline
PII   & 1152.8 & $25.4\pm5.1$ \\
\tableline
SII   & 1250.5 & $29.8\pm3.3$ \\
SII   & 1253.8 & $45.3\pm3.2$ \\
SII   & 1259.5 & $57.5\pm3.1$ \\
\tableline
ArI   & 1048.2 & $43.7\pm5.7$ \\
\tableline
FeII  & 2344.2& $163.7\pm15.6$\\
FeII  & 1608.4& $76.1\pm7.0$\\
FeII  & 1144.9& $60.9\pm8.6$\\
FeII  & 1121.9& $40.0\pm6.5$\\
FeII  & 1143.2& $30.1\pm5.9$\\
FeII  & 1125.4& $21.3\pm5.4$\\
FeII  & 2260.7& $19.4\pm4.8$\\
FeII  & 1133.6& $14.6\pm4.9$\\
\tableline
\tableline
\end{tabular}
\end{center}
\label{ews}
\end{table}

Constructing continuum fits to measure $F_c(v)$ for each line proved to be a tedious task due to mixing of the stellar spectrum and the cloud absorption spectrum.  Each line profile was fitted with a stellar absorption model, provided by S. J. O'Toole, prior to continuum fitting in order to identify stellar absorption lines. The stellar model, described in \citep{Otoole2004, Otoole2006}, was created using a metal line-blanketed LTE atmospheric model with solar metallicity and was synthesized using the LINFOR program. For each interstellar absorption line, first- to third- order Legendre polynomials were fitted to line-free neighboring regions within $\pm100 ~\kms$. 

Systematic errors in the column density arise from the chosen continuum fit and the chosen velocity range of the absorption line. The errors due to the chosen velocity range are estimated by varying the velocity range $\pm5~\kms$, while the errors associated with the continuum fit are found from the polynomial equation solver \citep[see ][]{SembachSavage1992}.  Figures \ref{spectra1} and \ref{spectra2} present each of the interstellar absorption lines along with its chosen continuum fit (blue line).  Fortunately, for all but three of the ionic species it was possible to fit continua that remained flat across the absorption line.  The \SiII-1304, \SII-1259, and \SII-1253 absorption lines were fit with higher order continua to account for stellar absorption indicated by the model of \citealt{Otoole2006}.  These three lines were also fit with flat continua to gauge the error from the higher order fits.  The average change of N$(v)$ for the \SiII-1304$\Angstrom$ and \SII-1259$\Angstrom$ absorption lines was an increase of 8\%.  The column density of the SII-1253$\Angstrom$ absorption line had a larger increase of 50\%, which was accounted for in its error value.

Line saturation poses the biggest source of error in the column density estimates.  In an unsaturated case all the lines of a single ion could be averaged to estimate the final column density.  However, a spread of N$_{\rm AOD}(v)$ values by at least a factor of two indicates possible saturation, and therefore a lower limit for the column density.  We found that there was a large spread of N$_{\rm AOD}(v)$ values between different absorption lines for the same ion.  In our AOD measurements, the largest spread of values ranged nearly two orders of magnitudes for \NI~ and \FeII, leading us to conclude that the AOD method was unreliable. 

Out of the eight ions detected, only \SII~, \FeII~, and \OI~ have the three or more absorption lines needed to adequately perform a $\chi^2$ test.  However, the \OI~ absorption lines are deep and affected by saturation and were therefore not used for the $\chi^2$ test.  The best fitting FWHMs were found to be $9.0^{+1.9}_{-4.0}~\kms$ for \FeII~ and $10.0^{+1.7}_{-2.7}~\kms$ for \SII.  Thus, a final mean FWHM of $9.5^{+1.8}_{-3.4}~\kms$ was used as a valid approximation of the FWHM for all ions, including the ones for which only one  line was measured.  For demonstration, Figure \ref{SIIcog} presents the $\chi^2$ minimization of the resolved \SII~ absorption lines, at 1250~$\Angstrom$, 1253~$\Angstrom$, and 1259~$\Angstrom$.  We find a column density for \SII~ of $\rm \log \{N_{cog}~[cm^{-2}]\}=14.71^{+0.08}_{-0.09}$. We note that we are effectively assuming that all ions reside in the same gas to convert the measured equivalent width to a column density.  Table \ref{colden} lists all the measured $\rm N_{cog}$ values for all eight ions detected. 

\begin{table}[!hb] 
\centering
\scriptsize
\caption{Column Densities and Metallicities}
\begin{tabular}{lccccc}
\tableline\tableline
Ion & log N$_{\rm cog}$ & Z\footnotemark & A\footnotemark &  $\rm [X/$\HI]\footnotemark \\
(1) & (2) & (3) & (4) & (5)\\
\tableline\tableline
NI    &      $15.37_{-0.68}^{+0.89} $ & $-4.67\pm0.89$ & -4.17 &  $-0.50\pm0.89$\\
OI    &      $15.88_{-0.57}^{+0.43} $  & $-4.16\pm0.43$ & -3.31 & $-0.85\pm0.43$\\
SiII  &     $14.89_{-0.23}^{+0.57} $   & $-5.15\pm0.57$ & -4.49 & $-0.66\pm0.57$\\
AlII  &     $13.37_{-0.64}^{+1.21} $   & $-6.67\pm1.21$ & -5.55 & $-1.12\pm1.21$\\
PII   &    $13.07_{-0.15}^{+0.16} $   & $-6.97\pm0.16$ & -6.59 & $-0.38\pm0.16$\\
SII   &    $14.73_{-0.10}^{+0.11} $   & $-5.31\pm0.12$ & -4.88 & $-0.43\pm0.12$\\
ArI   &    $13.53_{-0.20}^{+0.35} $   & $-6.51\pm0.35$ & -5.60 & $-0.91\pm0.35$\\
FeII  &   $14.25_{-0.17}^{+0.21} $   & $-5.79\pm0.21$ & -4.50 & $-1.29\pm0.21$\\
\tableline
\tableline
\label{colden}
\end{tabular}
\footnotetext[1]{Ion abundances measured from N$_{\rm cog}$ and $\log \{$ N(\HI) [cm$^{-2}]\}=20.04\pm0.04$}
\footnotetext[2]{Solar elemental abundances taken from \citet{Asplund2009}.}
\footnotetext[3]{Ion abundances with respect to the solar abundance values.}
\end{table}

\subsection{The \HI~ Column Density}  
We estimated the \HI~ column density in the direction of PG1144+615 using the nearest GBT 21-cm emission spectrum.  The top panel of Figure \ref{spectra1} shows the GBT spectrum closest to the background source.  It can be clearly seen that the cloud has two velocity components, a broad component at a velocity of $-51 \kms$ and a narrow component at $-48 \kms$.  The total column density, N(\HI), was estimated by integrating the brightness temperature $\rm T_B$ over the velocity interval of the two components, resulting in a total column density of $\rm \log \{N($\HI$)[\cm^{-2}]\}=20.13 \pm0.02$.  

\begin{figure*}[!hb]
\begin{center}
\includegraphics[width=5in]{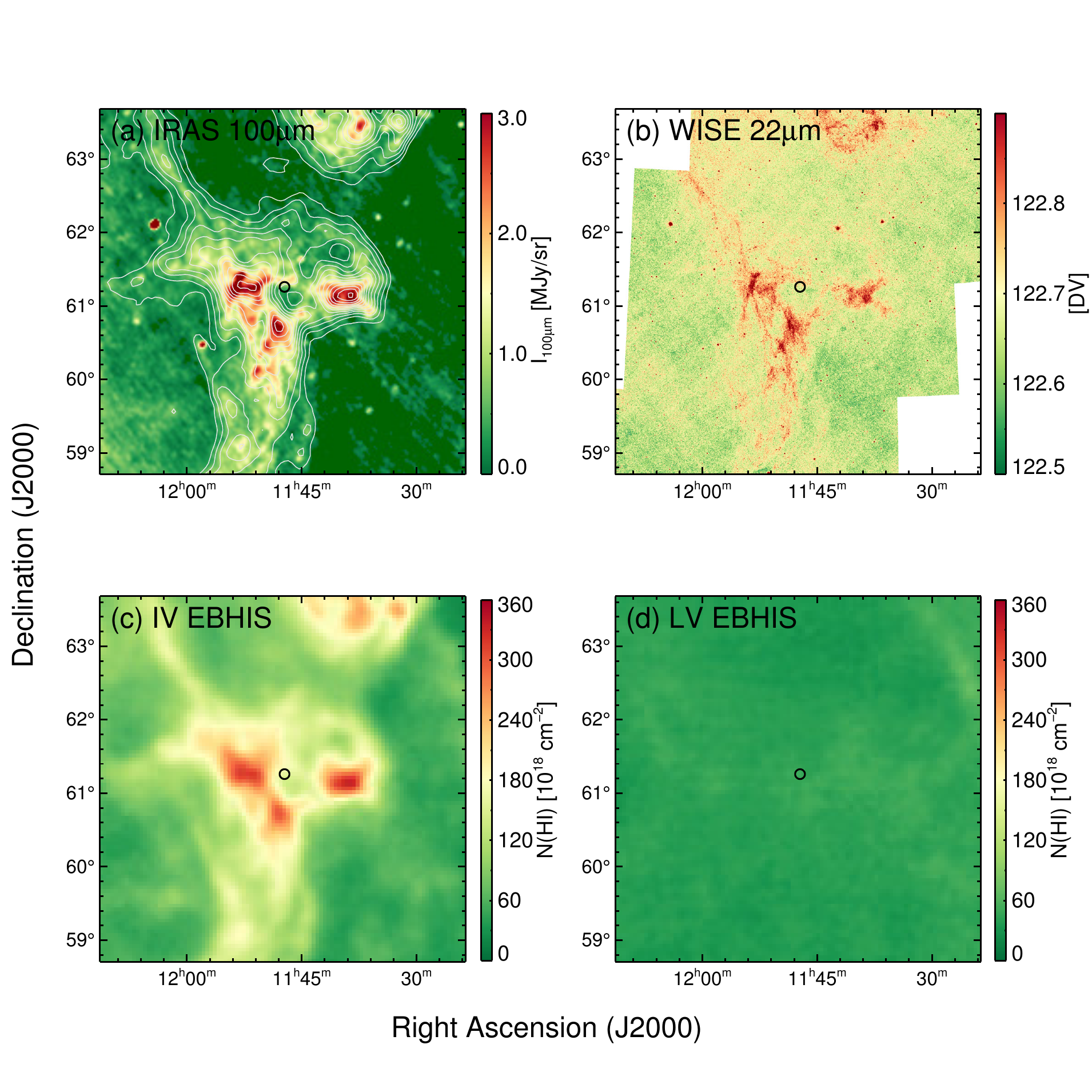} 
\end{center}
\caption{\small{Images of the IV21 cloud (5$^{\circ}\times5^{\circ}$).  The location of PG1144+615 is marked with the EBHIS beam size of $9\arcmin.4$, to scale (black circle). 
\textit{(a)}: IRAS 100~$\mu$m map with a resolution of $3\arcmin\times5\arcmin$.  Overlaid contours show the \HI~ column density distribution with level steps equivalent to $5\sigma$. The distribution of \HI~ is well correlated with regions of bright FIR emission indicating the presence of dust.  
\textit{(b)}: WISE 22~$\mu$m emission map with a resolution of $12\arcsec$.  A visual inspection revals no sign of  small-scale structure which would cause N(\HI) to be underestimated.   We also find no evidence of embedded point sources indicating active star formation.}
\textit{(c)}: \HI~ column density as measured from the EBHIS 21-cm emission map with resolution of $10\arcmin.7$. \HI~ emission was integrated over the velocity range of $-75~\kms$ to $-35~\kms$. Note, the EBHIS data was only used for an assessment of the cloud structure. 
\textit{(d)}: EBHIS \HI~ column density of the low-velocity gas component, integrated over the range of $-30~\kms$ to $30~\kms$. 
}
\label{grid1}
\end{figure*}

\begin{figure*}[!hb]
\begin{center}$
\begin{array}{c}
\includegraphics[width=6.5in]{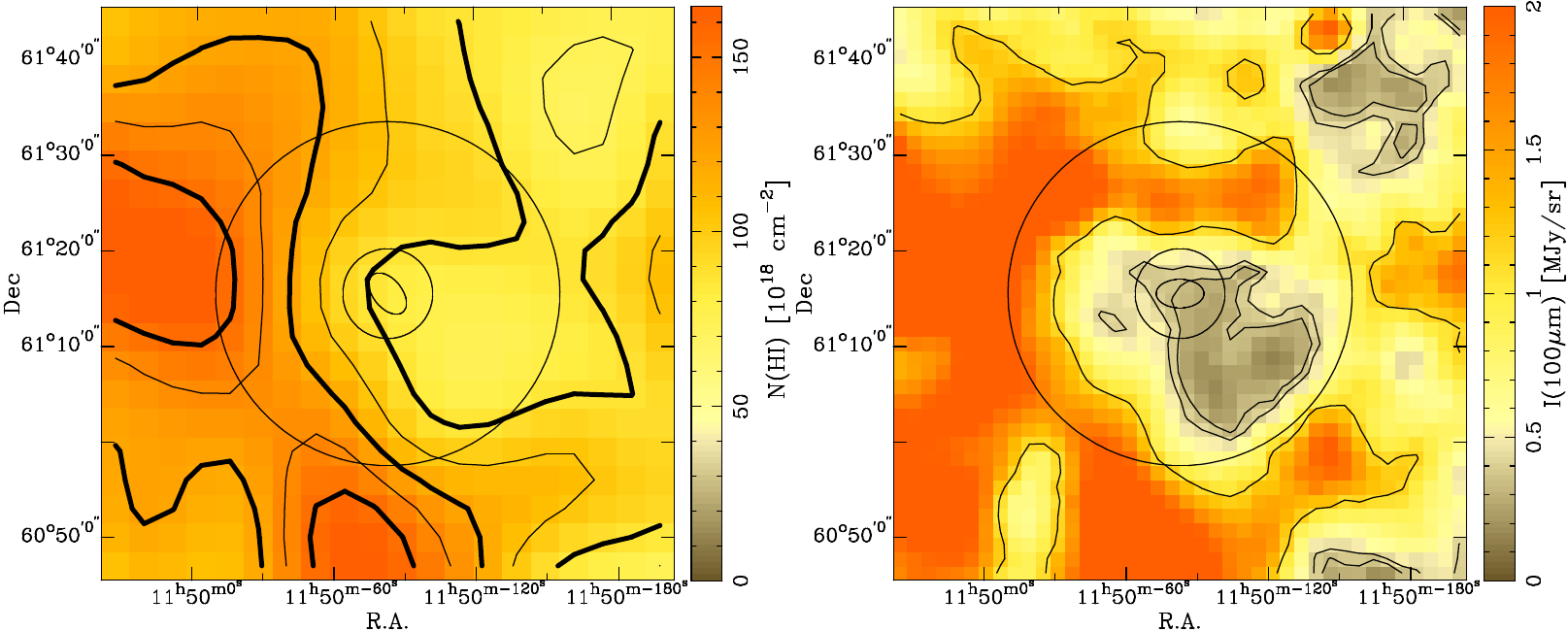} \\
\end{array}$
\end{center}
\caption{\small{Zoomed (1$^{\circ}\times1^{\circ}$) images of the IV21 cloud showing in detail the cloud structure near PG1144+615.  The location of PG1144+615 is marked with various beam sizes: the elliptical IRAS beam ($3\arcmin \times 5\arcmin$), the EBHIS beam ($9.4\arcmin$), and the LAB beam (36\arcmin).
 \textit{(a)}: \HI~ column density as measured from the EBHIS 21-cm emission map with pixel scale of $3\arcmin$. The levels are at 1.3, 1.9, 2.5 $\times 10^{20} \cm^{-2}$ (thin contours) and 1.6, 2.2, 2.8 $\times 10^{20} \cm^{-2}$ (thick contours), where $1.35\times10^{20} \cm^{-2}$ corresponds to the value seen toward PG1144+615 in the EBHIS beam.  
\textit{(b)}:  IRAS 100~$\mu$m map with a pixel scale of $1.5\arcmin$. The contour levels represent the intensity levels of the three beam sizes, 0.35 MJy/sr at 4\arcmin, 0.43 MJy/sr at 9.4\arcmin, and 1.14 MJy/sr at 36\arcmin.}
}
\label{grid2}
\end{figure*}

\begin{figure*}[!hb]
\begin{center}$
\begin{array}{c}
\includegraphics[width=4.8in]{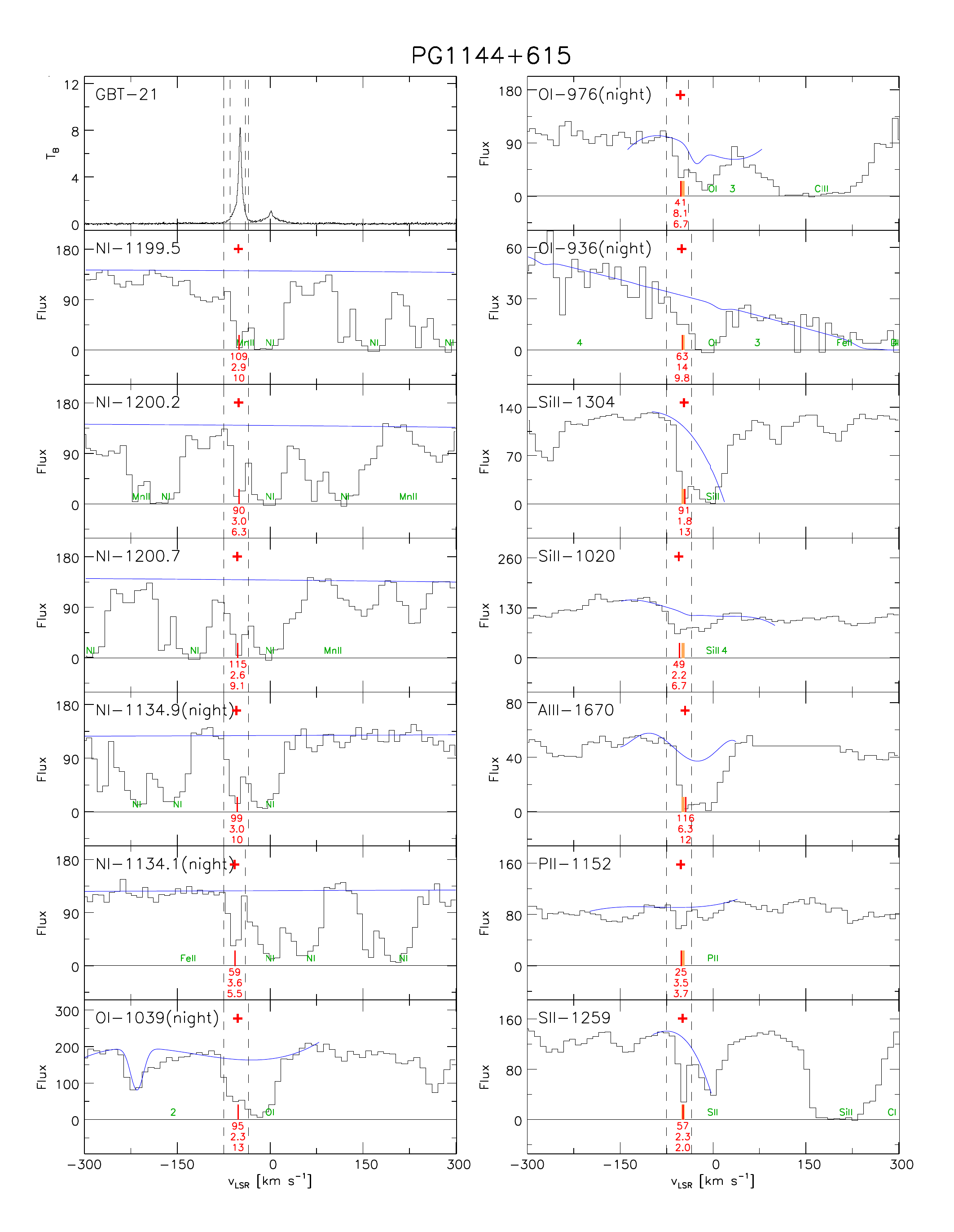} 
\end{array}$
\end{center}
\caption{\small{
\textit{Top Left:} The GBT 21-cm spectra in the direction of PG1144+615.  
\textit{Remaining Panels:} 13 rest-frame FUV absorption line spectrum detected for eight ions towards PG1144+615, black solid line, over the velocity range of $\pm300 ~\kms$.  The estimated continuum fit for each line is shown for each line by a blue solid line.  We note that the chosen continuum fits do not model the continuum for other absorption lines that lie adjacent to the IVMC absorption.  This divergence is due to using minimal order polynomials to fit only the IVMC absorption line and not those in the surrounding regions. These polynomial fits were made without the O'Toole stellar model because it often did not resemble the spectrum precisely.  Our attempt to account for stellar absorption lines can be seen from the unusual continuum shape for \SiII~1304 and the fact that the \SiII~1259 continuum near $-20~\kms$ does not go through the data. \textit{Cont. Fig. \ref{spectra2}}. }
}
\label{spectra1}
\end{figure*}

\begin{figure*}[!hb]
\begin{center}$
\begin{array}{c}
\includegraphics[width=5.2in]{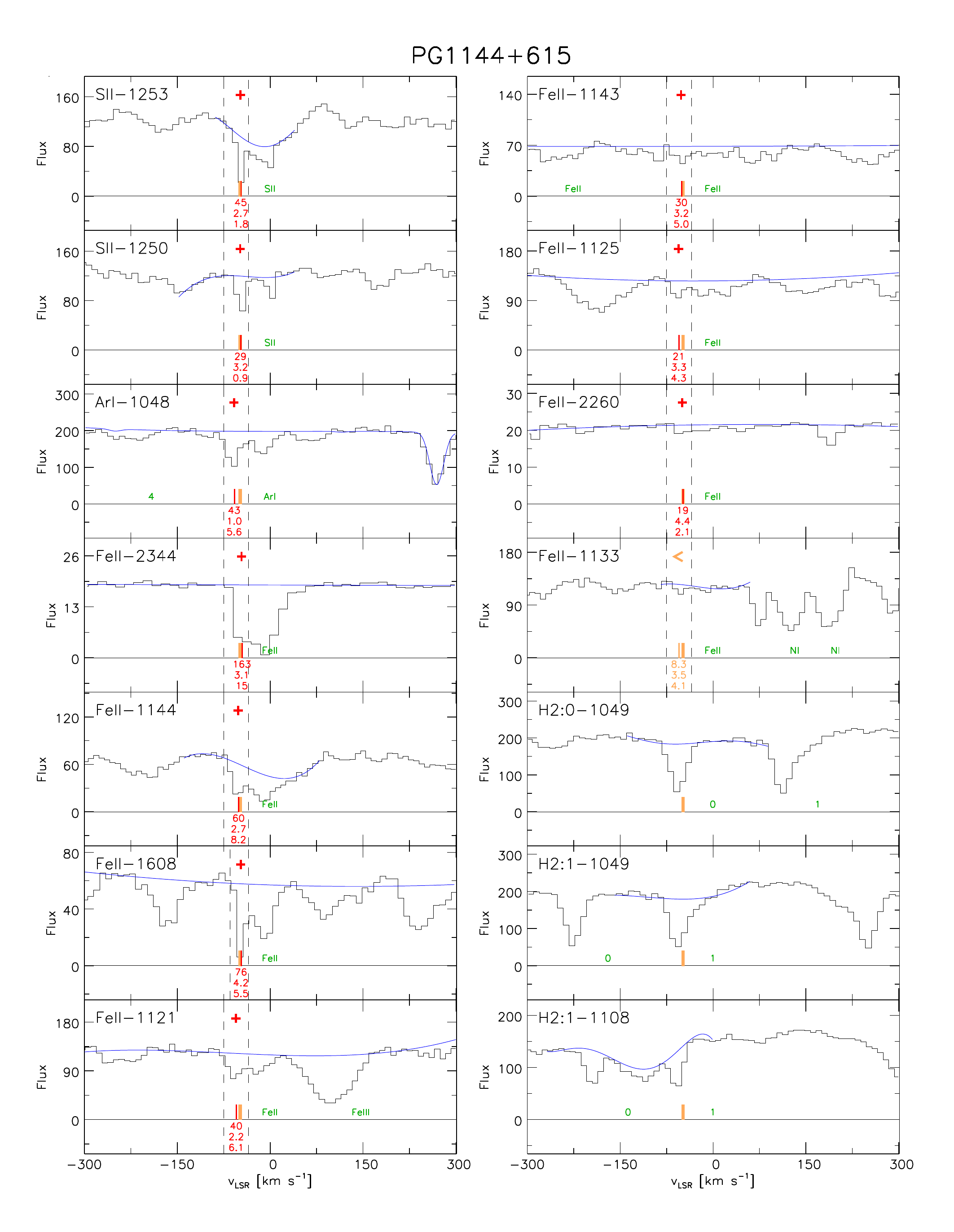} 
\end{array}$
\end{center}
\caption{\small{
The remaining eight rest-frame FUV absorption line spectra detected for eight ions towards PG1144+615 as described in Figure \ref{spectra1} .  The three left-hand bottom panels are examples of the FUV $\rm H_2$ absorption lines.  The chosen velocity interval of $dv=-75$ to $-35 ~\kms$ is expressed by the two dotted vertical lines. Noted within the velocity interval is the absorption line's measured equivalent width and uncertainty. Other interstellar absorption lines located in neighboring regions are noted in green text, where individual numbers refer to H$_2$ and its corresponding transition's upper J level.}
 }
\label{spectra2}
\end{figure*}

\begin{figure*}[!hb]
\begin{center}$
\begin{array}{c}
\includegraphics[width=6in]{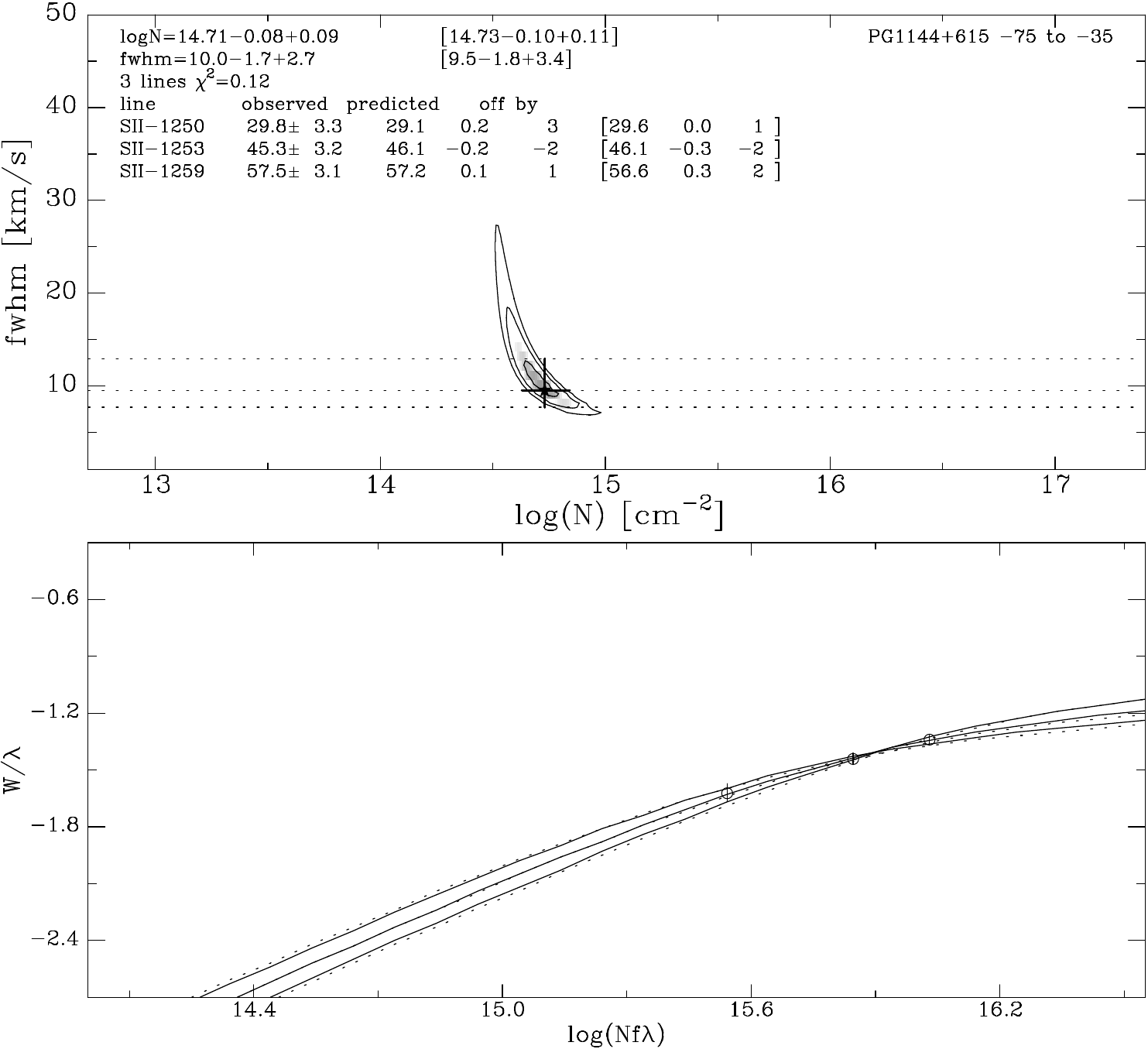}
\end{array}$
\end{center}
\caption{The SII curve-of-growth.  Solid lines give N for a FWHM$=10.0^{+1.7}_{-2.7}$, which is the best fit using only the SII lines.  Dotted lines are the curve-of-growth for FWHM$=9.5^{+1.8}_{-3.4}$, which is the final best FWHM using information from the the \FeII~ lines.
}
\label{SIIcog}
\end{figure*}

As indicated by the EBHIS maps, the widespread 21-cm emission is well correlated with far-infrared (FIR) emission from the IVMC, which traces dust heated by the interstellar radiation field (Figure \ref{grid1}).  Although bright FIR emission is unusual for most IVCs \citep[e.g.][]{Wakker2006}, its presence suggests that IV21 has a substantial dust component.  In Figure \ref{grid2} we present two $1^{\circ}\times1^{\circ}$ images of the IV21 region to show the cloud structure in the vicinity of PG1144+615.  The panel on the left shows the EBHIS N(\HI) map with a 9\arcmin.4~ resolution, while the right shows the IRAS 100~$\mu$m map with a resolution of 3\arcmin. We observe no significant sub-structures in the vicinity of PG1144+615 through a visual inspection of the 22~$\mu$m WISE image (Fig \ref{grid1}b). 
  
To quantify the effects of possible small-scale structure, we estimated the \HI~ column density by simulating a smaller \HI~ beam using the 100~$\mu$m IRAS images.  \citet{Boulanger1985} showed that the IRAS infrared dust emission of a galactic cirrus cloud is well correlated with its  \HI ~emission, with a consistent ratio between the mid-infrared emission and the \HI~ column density of $\rm I_{100 \mu m}$/N(\HI$)=1.4\pm0.3\times10^{-20}$~ $(\rm MJy~sr^{-1} ~cm^2)$ for a $20^{\circ}\times18^{\circ}$ high latitude field.  \citet{BoulangerPerault1988} performed a large-scale study of the infrared emission originating from different ISM components within 1 kpc of the Sun through comparisons of IRAS infrared maps, \HI, CO, and radio-continuum observations. Although their findings suggest that the dust abundance and interstellar radiation field (ISRF) are constant on scales on order of 100 pc, they found a factor of three spread in $\rm I_{100 \mu m}/$N(\HI) from one field to another.  For example, the Northern (b$>50^{\circ}$) and Southern (b$<-50^{\circ})$ Galactic caps were found to have $I_{100 \mu m}/$N(\HI) values of $0.92\pm0.14\times10^{-20}$~ $(\rm MJy~sr^{-1} ~cm^2)$ and  $0.79\pm0.06\times10^{-20}$~ $(\rm MJy~sr^{-1} ~cm^{2})$, respectively.  Conversely, fields between $|b|=30^{\circ}$ have a $\rm I_{100 \mu m}/$N(\HI) between $(1.1 - 1.4)\times10^{-20}$~ $(\rm MJy~sr^{-1} ~cm^{2})$, and for $|b|>10^{\circ}$, $\rm I_{100 \mu m}$/N(\HI$)=0.85\pm0.05\times10^{-20}$~ $(\rm MJy~sr^{-1} ~cm^{2})$.  They argued that this spread was due to variations in the ISRF, as the largest values are seen in the vicinity of OB associations, while the lowest values are in the direction of the Galactic anti-center.  

We use the IRAS 100 $\mu$m emission map of the cloud region to estimate the FIR flux of IV21 for a range of beam sizes, including that of the GBT. We did this by degrading the resolution of the FIR map using the IDL routine FILTER\_IMAGE to convolve from the IRAS beam size at 100~$\mu$m to the desired beam size.  To estimate the implied $I_{100 \mu m}/$N(\HI) at the resolution of the 21-cm data, we smoothed the 100~$\mu$m map to the resolution of the GBT telescope ($\theta_{FWHM}=9.1\arcmin$), finding the intensity in the direction of PG1144+615 to be $\rm I=0.43~(MJy~sr^{-1})$.  We find an intensity to column density ratio $I_{100 \mu m}/$N(\HI) of $0.32\times10^{-20} (\rm MJy ~sr^{-1} ~cm^2)$ using the measured $\rm \log \{N($\HI$)[\rm cm^{-2}]\}$ of 20.13. If taken at face value, this suggests an FIR to N(\HI) ratio in IV21 that is only one-third the usual ratio for the solar neighborhood of $\sim1\times 10^{-20} \rm ~(MJy~sr^{-1}~cm^2)$.  

Using the $I_{100 \mu m}/$N(\HI)  ratio we estimated the implied N(\HI) as a function of resolution for various beam sizes (Table \ref{Nest}).  The FIR data provides some evidence for small-scale structure inside the 9\arcmin.1 GBT beam as at the full 4\arcmin~ IRAS 100 $\mu$m resolution, the FIR emission is 74\% of the value found when smoothing to 9\arcmin.1.  However if the cloud's abundance were solar, this would imply that N(\HI) is five times smaller when measured in a pencil-beam toward PG1144+615 as compared to the measurement in the 9\arcmin.1 GBT beam.  This is rather unlikely.  For example, \citet{Wakker2001b} studied the differences between \HI~ column densities measured from various beam sizes using high resolution (1\arcmin-2\arcmin) \HI~ data of 8 HVC/IVC probes with 21-cm interferometric data (Westerbork telescope and ATCA).  In a comparison to \HI~ column densities estimated from lower resolution observations made with the Effelsberg telescope they found that the N(\HI; 1\arcmin-2\arcmin)/N(\HI; 9\arcmin) ratio varies by less than 25\%.  More recently, in a study of the small scale structure of the ISM, \citet{Wakker2011} used Ly$\alpha$ absorption line data from STIS to measure the \HI~ column density in a pencil beam toward 59 AGN for comparison with column densities measured from radio observations (9\arcmin - 36\arcmin; GBT and LAB survey).  They found that on average the N(\HI; Ly$\alpha$)/N(\HI; 21-cm) ratio varies between 0.8 and 1.2.  Effectively, \citet{Wakker2001b, Wakker2011} found that a 20\% systematic error is needed to account for small scale structure when measuring N(\HI) with a 9\arcmin~ beam, and that this error decreases with decreasing beam size. To account for our beam correction and the possibility of still unaccounted for small scale structure, we include a systematic error in N(\HI) for the calculation of the abundance.  Since we simulate the equivalent of a 4\arcmin~ beam, we use a 10\% systematic error, making our final estimate $\rm \log\{$N(\HI)$[\cm^{-2}]\}=20.04\pm0.04$.  We note that the direction of PG1144+615 is unusual in that the total N(\HI) is dominated by the IVC \citep[see][]{LockmanCondon2005}.

\begin{table}[!hb] 
\centering
\scriptsize
\caption{N(\HI) Implied by I(100$\mu$m) as a Function of Beam Size}
\begin{tabular}{lcccc}
\tableline\tableline
$\theta^{\prime}$\footnotemark  & I$_{100\mu m}$\footnotemark & $\rm I/I_{9\arcmin.1}$\footnotemark & $\rm \log$ N(\HI)\footnotemark & $\rm \log$ N(\HI)$_{\rm obs}$\footnotemark  \\
\tableline\tableline
    4.0\tablenotemark  &   0.35  & 0.74 &20.04 &\nodata \\	
    6.0  				&   0.34  & 0.80 &20.03 &\nodata \\  
    8.0  				&   0.39  & 0.92 &20.09 &\nodata \\
    9.1  				&   0.43   &1.00  &20.13 &20.13 (GBT) \\
    9.4  				&   0.44  & 1.02 &20.14  &20.20 (EBHIS) \\  
   15.0 				&   0.64  & 1.49 &20.30 &\nodata \\  
   20.0 				&   0.81  & 1.89 &20.41  &\nodata \\ 
   36.0				&  1.14   & 2.66 &20.55 & 20.31(LAB) \\
 \tableline
 \tableline
\label{Nest}
\end{tabular}
\footnotetext[1]{Beam size}
\footnotetext[2]{The measured IRAS 100 $\mu$m intensity $\rm (M~Jy~sr^{-1})$ for $\theta^{\prime}$.}
\footnotetext[3]{The ratio of the 100 $\mu$m intensities between $\theta^{\prime}$ and $\theta =9\arcmin.1$ (GBT).}
\footnotetext[4]{The HI column density implied by I$_{100\mu m}$ and $I_{100 \mu m}/$N(\HI)$=0.32\times10^{-20} (\rm MJy ~sr^{-1} ~cm^2)$.}
\footnotetext[5]{The observed HI~ column density measured from the original \HI~ data map.}
\footnotetext[6]{I$_{100\mu m}$ in the original IRAS  beam ($3\arcmin \times 5\arcmin$).}
\end{table}

We note that our estimate of the $\rm I_{100 \mu m}/$N(\HI) ratio is also lower than that estimated for the two clumps bracketing PG1144+615 by \citet{Weiss1999}. They found $\rm I_{100 \mu m}/$N(\HI)  to be $\rm 0.93\pm0.18\times10^{-20} ~(MJy~sr^{-1} ~cm^2)$ and $\rm 0.97\pm0.20\times10^{-20} ~(MJy~sr^{-1} ~cm^2)$ for the left and right clumps, respectively.  However, these estimates were averages for each clump and their surrounding regions.

There is a faint low-velocity gas component ($\pm30~\kms$, Fig. \ref{spectra1}), which is widespread across the whole IVMC region.   Unlike the 21-cm spectrum, where the IVMC and low-velocity gas components can be separated in velocity space, the infrared data displays a composite of the dust emission from both components. Thus, the FIR emission (Fig \ref{grid1}a) will be contaminated by the emission of the low-velocity component.  Ideally, the total 100~$\mu$m intensity should be equal to the sum $j_{LVC}$N(\HI,L)$+j_{IMVC}$N(\HI), where N(\HI,L) is the \HI~ column density of the low velocity component, and $j_{LVC}$ and $j_{IMVC}$ are the ratios between the FIR emission and the HI column densities for both components.   To estimate the contribution of $j_{LVC}$N(\HI,L) to the total 100~$\mu$m intensity, we performed a $\chi^2$ test using the measured I$_{100\mu m}$ of $0.43~(\rm MJy~sr^{-1})$ and the estimated column densities of the IVMC and low-velocity gas from the GBT data.  We found that while $j_{IMVC}$ behaved fairly well and can be constrained to values $\pm30\%$ of the measured $I_{100 \mu m}/$N(\HI) ratio of $0.32\times10^{-20} ~(\rm MJy ~sr^{-1} ~cm^2)$, $j_{LVC}$ failed to converge to a physical value.  We interpret this finding as an indication that the contribution of the low-velocity gas to the total I$_{100\mu m}$ is negligible. 
  
\subsection{Metallicity Estimates}
We measured the ion abundances, defined as $\rm Z\equiv \log (X/$\HI), for each ion, X, using line ratios of the ion column densities, N$_{\rm cog}$, and N(\HI).  The total error on the abundance was estimated by standard linear error propagation of the total errors in N$_{\rm cog}$ and N(\HI). For each abundance estimate the total error is expressed in terms of upper and lower errors due to their logarithmic conversion. All the estimated ion abundances and respective errors are presented in Table \ref{colden}.

To measure the metallicity of IV21, $\rm Z/Z_{\odot}$, we first calculated the individual ion abundances with respect to solar abundances, $\rm[X/$\HI$]\equiv \rm Z - A$, where A is the logarithmic standard solar abundance estimates taken from the solar composition review of \citet{Asplund2009}.  All of the measured ion abundances with respect to solar values are presented in Table \ref{colden}.  The most reliable measurements are for \SII~ and \FeII, which were the only ions with three or more absorption lines need to constrain the column densities using the curve-of-growth method.  In principle, the [\SII/\HI] ratio can differ from S/H because of ionization effects (i.e. part of the cloud may be ionized containing both \HII, \SII, and \SIII).  However, this effect is unimportant at \HI~ column densities above $\sim5\times10^{19}~\rm cm^{-2}$.  Also, in principle, the [\OI/\HI] ratio would give a better metallicity than [\SII/\HI], since the ionization of \OI~ is coupled to that of \HI~ through a charge-exchange reaction.  However, the relatively small FWHM-value coupled with the strength of the measurable \OI~ lines makes the derived \OI~ column density rather uncertain. Finally, due to the large dust content of the cloud, the \FeII~ metallicity most likely does not accurately describe the overall cloud metallicity.  Thus, using the \SII~ column density only, we find that IV21 has a sub-solar metallicity of $\rm \log (Z/Z_{\odot})=-0.43\pm0.12$ dex.   

\begin{figure*}[!hb]
\begin{center}$
\begin{array}{c}
\includegraphics[width=5.5in]{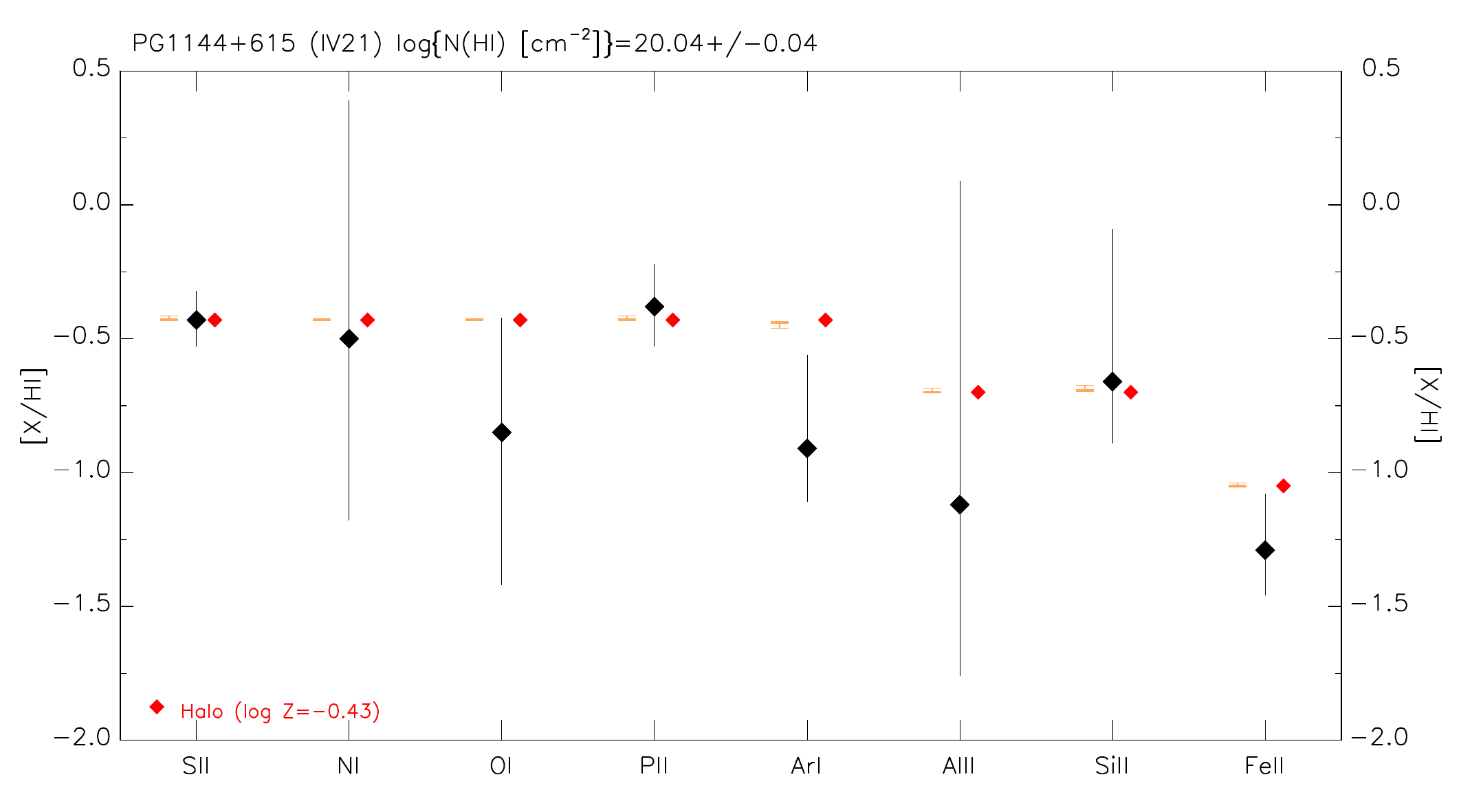}
\end{array}$
\end{center}
\caption{\small{The estimated ion abundances, with respect to solar values, for IV21, [X/\HI].  Our estimated ion abundances from FUV spectra are shown by the black diamonds.  The red diamonds show the expected values for a standard halo depletion pattern, assuming an overall cloud metallicity of $\rm \log (Z/Z_{\odot})=-0.43$ dex.  The ranges shown by the orange bars indicate the spread of ionization corrections found using the CLOUDY models assuming a fixed distance of 1 kpc and a range of densities: $\rm n=1 ~\rm cm^{-2}$ (thick horizontal bar) and $\rm n=10 ~\rm cm^{-2}$ (thin horizontal bar).  We find that our results are well correlated with those predicted by the CLOUDY photoionization models.}
}
\label{abd}
\end{figure*}

To examine the effect of ionization structure on the ion abundances we used the photoionization code CLOUDY \citep{Ferland1998} to model a sheet-like IVC of uniform density.  We used the simplest cloud geometry since the overall three-dimensional structure of the IVC is unknown and to minimize computing time.  Due to the location of IV21, we set the CLOUDY abundances to the halo depletion pattern of \citet{SavageSembach1991} and assumed a cloud metallicity of $\rm \log (Z/Z_{\odot})=-0.43$~dex.  We used the interstellar radiation field (ISRF) described in \citet{Fox2005b}.  This model combines a spectrum ranging from 90 to 912 $\Angstrom$ (dominated by OB stars) taken from \citet{BlandHawthorn1999} with the solar neighborhood spectrum of  \citet{Mezger1982} from $912-2400~\Angstrom$.  The derived metallicities and errors for all eight ions are displayed in Figure \ref{abd} (black points).  For comparison, the ion abundances implied for a cloud with a metallicity of $-0.43$ solar and a standard halo-like depletion pattern are also displayed (red points).  A range of expected metallicities was predicted for a spread of ionization corrections given a fixed distance of 1 kpc and a range of volume densities ($\rm n=1$ and $10 ~\rm cm^{-3}$).  These expected metallicity ranges are also shown in Figure \ref{abd} as orange brackets and listed in Table \ref{colden}.  All of the measured metallicities in IV21 are in agreement with the expected values for a cloud with $\rm \log (Z/Z_{\odot})=-0.43$~dex, a halo-like depletion pattern and CLOUDY-based ionization corrections given the assumed radiation field.  However, predicted \FeII, \ArI, and \AlII~ metallicities do fall slightly below their expected values. The large uncertainty for \OI~ is due to the fact that its absorption lines are deep and affected by saturation.

\subsection{Determination of \htwo~Parameters}
Following the methods described in \citet{Wakker2006} we derived the \htwo~column density for each rotational (J) level in the direction of PG1144+615 using the FUV absorption spectra.  This method includeds several steps, beginning with the continuum fitting of the various selected \htwo~absorption lines.  The underlying stellar spectrum complicated the continuum fitting as many \htwo~ lines were seen against  stellar absorption lines. Thus, individual continuum fits were made for each \htwo~ absorption line in the same manner as for the ion absorption lines (See Section 3.1).

Next, we selected the \htwo~lines which were deemed useful for measuring accurate equivalent widths.  Beginning with the normalized spectrum, sets of lines ordered by oscillation strength for each J level were displayed.   Through visual inspection, absorption lines were chosen if they did not contain : (1) interstellar metal lines, (2) neighboring \htwo~lines, (3) intervening intergalactic absorption, intrinsic AGN lines, and geocoronal emission. For each (J) level, we measured the equivalent width of each \htwo~absorption line over a velocity range of $\sim 25 ~\kms$ centered around $-48~\kms$.  Column densities and FWHMs were estimated for each (J) level using a $\chi^2$ minimization test (Equation \ref{chi}).  

The estimated column density depends strongly on the estimated FWHM.  The best fit FWHM for each J level will generally be different due to random noise and uncertainties in the fitted continuum.  It is also possible that the widths differ intrinsically, but our data do not allow us to test for that possibility.  However, if we assume that all the J-levels have the same FWHM, the best resulting column density is found through the minimization of $\chi^2$ given a fixed value of the FWHM.  

From the resulting column densities, excitation temperatures were estimated through the Boltzmann distribution:
\begin{equation}
\rm \frac{N(J+1)}{N(J)}=\frac{g_{J+1}}{g_J}\exp\left(\frac{-(E_{J+1} - E_J)}{kT_{J(J+1)}}\right),
\end{equation}
where $g_J$ are the statistical weights and E$_J$ are the excitation energies of the different levels. Thus, for the estimated column densities for the first three rotational levels the three column density ratios $\rm N(1)/N(0)$ and $\rm N(2)/N(1)$ are solved for the temperatures T$_{01}$ and T$_{12}$.  Theoretically, these two temperatures will be equal if the \htwo~is purely collisionally excited and in equilibrium. However, due to radiative excitation it is generally observed that T increases with increasing J-level. 

Through an analysis of the first four J-levels we estimated the molecular hydrogen column density of the cloud to be $\rm \log \{N(H_2)[cm^{-2}]\}=16.72^{+0.17}_{-0.27}$.  Additionally, we find a mean FWHM of $W=6.5\pm0.6 ~\kms$ with associated excitation temperatures of  $T_{01}=148\pm21$ K and $T_{12}=T_{32}=171\pm15$ K.  The lower panels of Figure \ref{spectra2} presents selected \htwo~spectra along with their estimated continua.

\section{Discussion}
Our metallicity estimate of the IVMC IV21 is based on 23 interstellar absorption line of eight ions combined with 21-cm line observations.  All of the measured ion abundances indicate a sub-solar metallicity for this cloud.  Our best estimate of the metallicity of the cloud, including the ionization correction and halo depletion patterns is $\rm \log (Z/Z_{\odot})=-0.43\pm0.12$~dex, assuming the solar abundances of \citet{Asplund2009}.  This measurement of a sub-solar metallicity is consistent with the cloud's anomalous value of the I$_{100\mu m}$/N(\HI) ratio, which is only one-third of the usual solar-neighborhood ratio of $\sim 1\times10^{-20} \rm ~(MJy~sr^{-1}~\cm^{2})$ \citep[e.g.][]{BoulangerPerault1988, Weiss1999} \textit{This makes IV21 the only known cloud in the solar neighborhood to have a significantly sub-solar metallicity.} In this section, we discuss the possible cause for this and some of the possible implications. 

\subsection{The Origins of a Sub-Solar Metallicity for IV21}
One caveat for our claim that IV21 has a sub-solar metallicity is that our \HI~ reference column density used to determine the metallicity is obtained by comparing the ``pencil-beam'' line-of-sight for FUSE/HST to a broader, simulated 4\arcmin~ beam obtained with the GBT and IRAS data.  If there is a factor-of-five variation in the \HI~ column density within the 9\arcmin.1 beam of the GBT, the metallicity of this cloud could be consistent with solar. As discussed in Section 3.2.2, we consider this possibility to be unlikely, both because of the relative smoothness of the diffuse infrared emission at sub-\HI~ beam scales as well as the fact that a systematic comparison of beam size effects in many other directions only indicates a 20\% variation in the \HI~ column density within a $10\arcmin$ beam.

If the sub-solar metallicity is therefore real, how did such a low-metallicity molecular cloud come to be in the solar neighborhood? There are two reservoirs of low metallicity gas associated with the Galaxy: the extreme outer disk and the extra-galactic/circum-galactic infall identified as HVCs.   A recent study of HII regions in the Galaxy indicates a metallicity gradient of $-0.03$ to $-0.07 \rm ~dex/\kpc$, implying that gas with Z/Z$_{\odot}\sim-0.5$ dex is typically at galactocentric radii of $\sim20$ pc \citep{Balser2011}. It is difficult to imagine a scenario in which low metallicity gas from the outer Galaxy has migrated 10 kpc inward in radius to the solar neighborhood while staying at $\rm z<500$ pc. This leaves the extra- or circum-galactic infall model as our preferred hypothesis. We will call this the ``HVC impact origin".

\begin{figure*}[!hb]
\begin{center}$
\begin{array}{c}
\includegraphics[width=7in, angle=0, trim=0 6in 0 0]{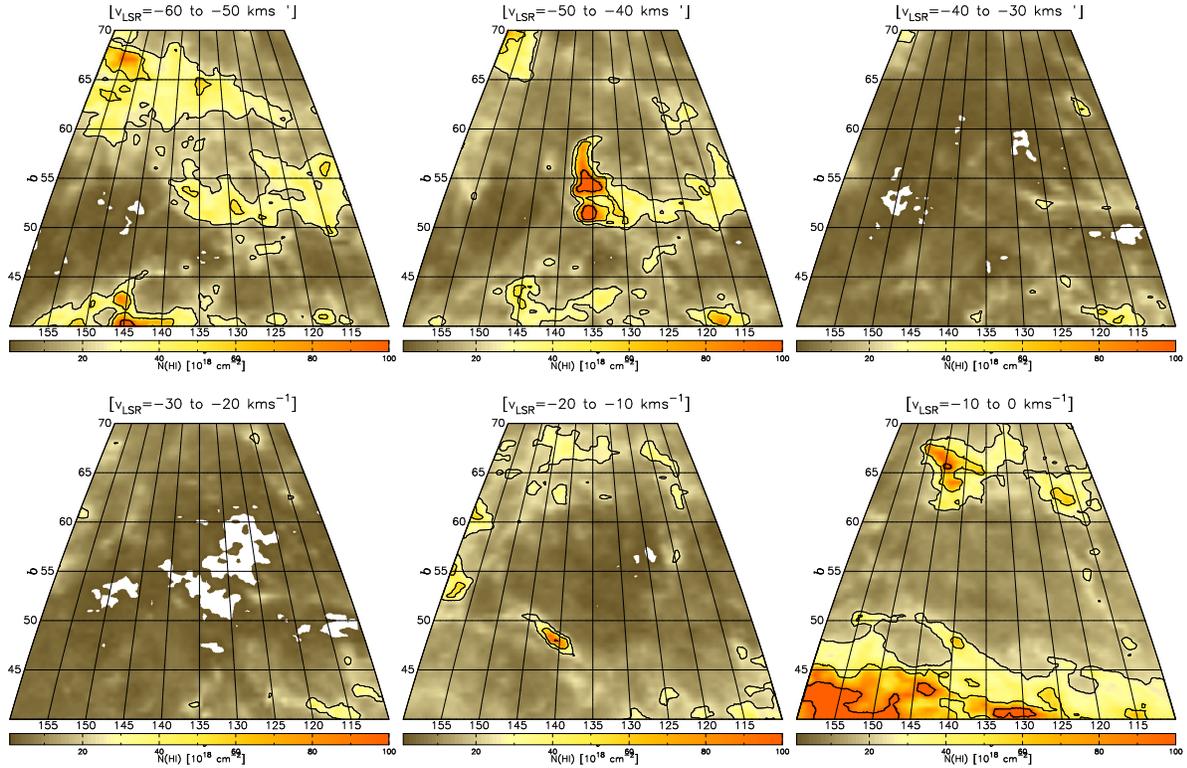}
\end{array}$
\end{center}
\caption{ IV21 extended region channel maps of \HI~ from the LAB survey.  The intensity maps are integrated from $-60~\kms$ to $0~\kms$ in steps of $10~\kms$. The upper-middle panel clearly displays IV21 at $v_{\rm LSR}\sim45~\kms$.  The lower panels ($v_{\rm LSR}>-30~\kms$) show a lack of halo gas at lower velocities, indicating that it may have been swept up by the metal-poor progenitor of IV21.}
\label{LABchan}
\end{figure*}

In addition to being the first known Galactic CO cloud with sub-solar abundance, IV21 is the only IVMC/IVC with a firmly-established sub-solar abundance \citep[e.g.][]{Richter2001a,Richter2001b,Sembach2004}. On the other hand, HVCs generally have sub-solar metallicities ranging from a low of $\sim 10\%$ of the solar value to $20-60\%$ of the solar value (see Table 2, Wakker \& van Woerden 2013).  IV21 appears to be in the middle of this metallicity range. Given the cloud's presence in the low-Galactic halo, it is possible that some mixing between the local metal-enriched ISM and a lower metallicity progenitor gas could have enriched this cloud, a possibility supported by the depletion seen in \FeII~ discussed in Section 3.3. A lack of low-velocity gas is revealed by the channel maps of IV21 and its surrounding regions (Fig. \ref{LABchan}).  This ``hole'' in the low-velocity gas has been known since the work of \citet{WesseliusFejes1973} and has never been adequately explained.  Our results suggest that it is possible that metal-enriched gas near the Galactic plane may have been swept up by the lower metallicity progenitor.  The potential for the mixing of different metallicity gas will complicate the interpretation of the metallicity and the dust emission, but would not invalidate our conclusion of a sub-solar abundance unless the ionization fractions are significantly out of equilibrium. 

Another piece of evidence in favor of a ``HVC impact origin'' for the cloud is, of course, the unusual kinematics of this object, even for the relatively sparse class of high-latitude CO clouds.  With a line of sight velocity of $-48 ~\kms$ at b=54.2$^{\circ}$, this cloud has a significant component of its motion towards the Galactic plane. An analysis of the gradient of the LSR velocity across the sky could put further constraints on the three-dimensional space velocity of this cloud, shedding further light on its origin \citep[e.g.][]{Lockman2008}. 

To summarize, we believe that the metallicity, unusual kinematics, anomalous I$_{100\mu m}$/N(\HI) ratio, and abundance patterns all point to an extra-galactic origin for this cloud.  How unique is this situation?   Although this is the first Galactic CO cloud known to be sub-solar, there are very few high-latitude molecular clouds for which we can obtain similar constraints on metallicity due to the lack of background targets.  The only other example of a suitable spectroscopic candidate is the sub-dwarf star PG1154-070 ($\rm d=1.3\pm0.3~\kpc$) which is located along a gaseous ``bridge'' joining the IVMCs IV288+54 ($v_{LRS}\sim -24~\kms$) and IV283+54 ($v_{LRS}\sim -34~\kms$).  With only eight known members, IVMCs are similar to IVCs in that they possess velocities which deviate by more than $20~\kms$ from values expected from simple models of Galactic rotation \citep{MagnaniSmith2010}.  Although, IVMCs differ from other IVCs shown to contain \htwo, such as those analyzed by \citet{Richter2003} and \citet{Wakker2006} which are primarily atomic with N(\htwo)$\rm <10^{17} ~cm^{-2}$. In contrast, the IVMCs contain \htwo~ column densities of $\rm \sim 10^{20} ~cm^{-2}$ as shown by the detection of CO within some of their densest cores \citep[e.g.][]{Weiss1999,MagnaniSmith2010}.  If we could measure the metallicity of IV288+53 and IV283+54 towards PG1154-070 and compare to the values estimated here for IV21, we would be able to determine if (a) IVMCs are a distinct set of clouds which represent the mixing of infalling low-metallicity gas with the lower Galactic halo, or (b) although all IVMCs have peculiar velocities and dense molecular gas, they originate from different gas sources leading to a range of metallicities. 

\subsection{The Implications of Sub-Solar Metallicity for IV21} 
With the caveats that (1) an unknown fraction of the mass associated with IV21 may be swept-up material with local ISM abundances and (2) we are probing the metallicity in a direction not directly aligned with the CO cores of this complex, this cloud presents an ideal nearby laboratory to study the process of molecular formation in a low-metallicity environment. 

\subsection{Extragalactic Infall vs. The Galactic Fountain}
Generally, IVCs are believed to represent the cool, return flows of the Galactic fountain.  A recent Galactic fountain model by \cite{Marasco2012} was used in an attempt to reproduce the Milky Way's \HI~ emission.  Their simulations modeled relatively cool gas clouds which first are ejected by supernovae, then ballistically orbited above the disk while interacting with coronal gas through ram-pressure drag and accretion which cools the cloud. In a comparison of their results with LAB \HI~ emission maps they found that their models reproduce the \HI~ emission at intermediate velocities, but not that seen at high-velocities.  However, it is unclear as to whether their results reproduce the emission seen in the IV-Arch \citep[see Fig. 10][]{Marasco2012}.  Many of the IVC systems that agree with their results are located $\lesssim3$~kpc away and contain near disk-like metallicities \citep[e.g.][]{vanWoerden2004}.  Although they note that IVCs should be somewhat metal-poor in the model since they are a mixture containing up to $24\%$ metal-poor halo gas.  Their models also predict an \HI~ accretion rate, produced by the cooling of coronal gas through interaction with fountain clouds, of $\rm \sim2 ~M_{\odot}~yr^{-1}$.  This is the rate needed to sustain the current rate of Galactic star formation. 

\begin{figure*}[!hb]
\begin{center}$
\begin{array}{c}
\includegraphics[width=4.5in, angle=0]{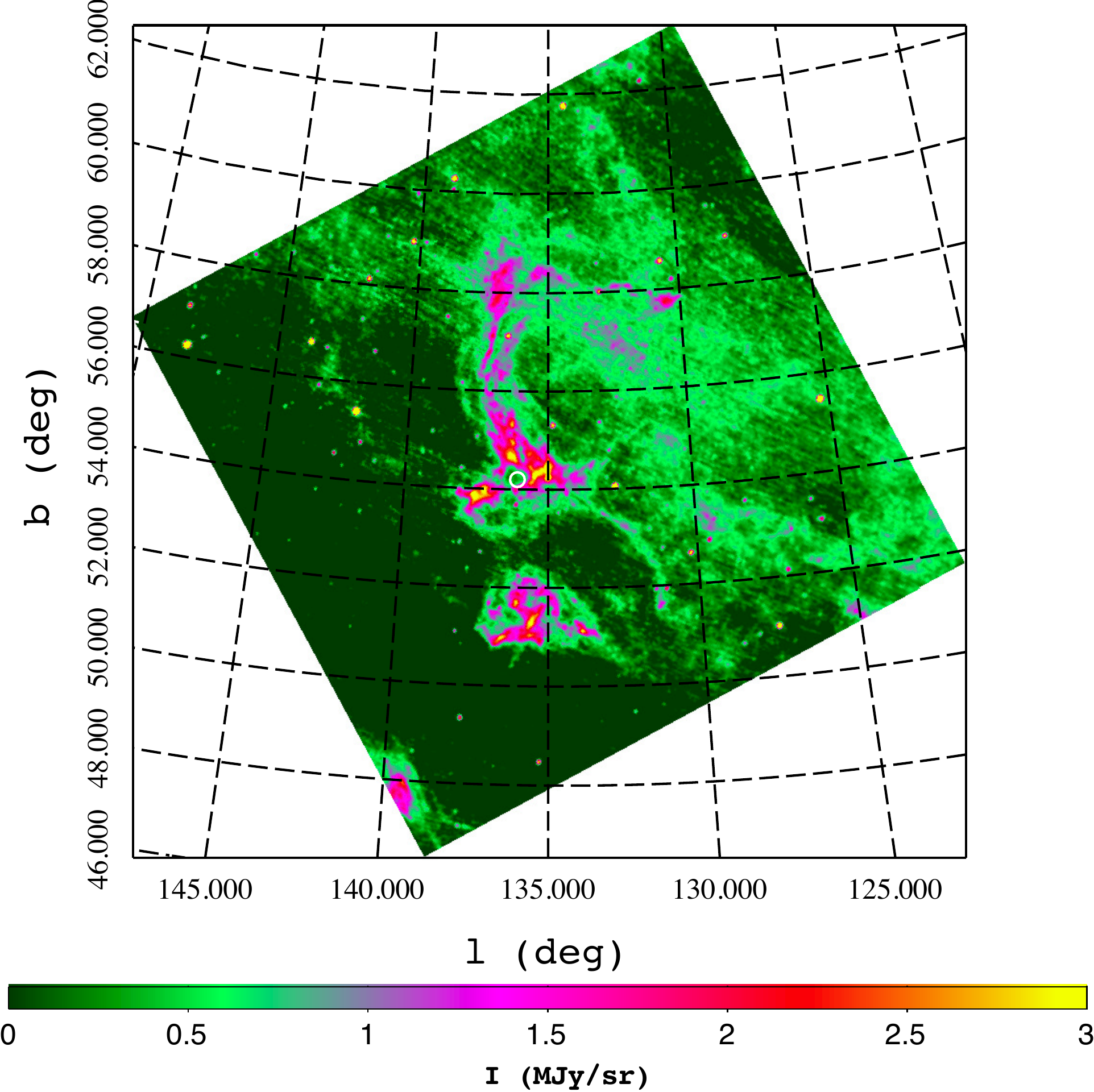}
\end{array}$
\end{center}
\caption{IRAS 100~$\mu$m image of IV21 with a 12$^{\circ}\times12^{\circ}$ bounding box and pixel scale of $1.5\arcmin$.  This image is a larger version of the map used in Figure \ref{grid1}a, but has been reoriented in galactic coordinates to demonstrate IV21's downward pointing, head-tail like cloud structure.  This cloud morphology is similar to the infalling HVC simulations described by \citet{HeitschPutman2009} }
\label{iras12}
\end{figure*}

Although located within $\sim3$~kpc of the solar neighborhood and in the direction of the IV-Arch ($>8$~kpc, near-solar metallicity), IV21 differs in distance and metallicity.  From IV21's sub-solar metallicity, we believe that it is the result of an infalling HVC mixing with the metal-enriched lower halo.  With a metallicity of $-0.43$ dex, IV21 is well below the metal-poor metallicities estimated by \cite{Marasco2012}.  The hydrodynamical simulations of \citet{HeitschPutman2009} have shown that it is possible for warm ionized gas stripped from an infalling HVC to cool and re-form as cold \HI~ clouds at intermediate velocities.  These IVCs will most likely be embedded in the warm ionized medium (WIM) at $\sim3$ kpc above the disk.  HVCs formed by infalling clouds with original masses of $\rm >10^{4.5}~M_{\odot}$ will retain less than 10$\%$~of their original \HI ~content, while cloud fragments that survive to lower velocities within the WIM will contain even less. The estimated total \HI~ mass of IV21 is 240 M$_{\odot}$ \citep{Weiss1999}, only 1$\%$ of the minimum mass of an infalling HVC. Thus, IV21 may not be the final remnants of a whole infalling HVC, but rather the result of gas stripped away from a much larger infalling cloud \citep{HeitschPutman2009}.  They note that the only way to distinguish between ``HVC impact origin'' IVCs and those formed by the Galactic fountain is through metallicity estimates.  The IVCs formed from infalling HVCs would have significantly sub-solar metallicities and head-tail like cloud structures. Figure \ref{iras12} presents a large scale $12^{\circ}\times12^{\circ}$ 100$\mu$m image from IRAS (reoriented in Galactic coordinates) which shows that IV21 has a tail-like counterpart streaming out to the north, consistent with the clouds modeled by \citet{HeitschPutman2009}.

\subsection{Dust and the Formation of \htwo}
Even with a sub-solar metallicity, IV21 has widespread FIR emission which closely agrees with the EBHIS \HI~ distribution (See Fig. \ref{grid1}a) and suggests that there is a substantial dust content throughout the cloud.  The origin of this large dust content still remains unclear.  The first possibility is that the dust formed within the cloud as it approached the Galactic disk.  However, it is difficult to explain how such a large amount of dust could form within a metal-poor cloud.  Another possibility is that dust which already resided in the lower Galactic halo, deposited by expanding supernovae shells, was swept up and mixed with the low metallicity progenitor as it approached the disk.  Lastly, the bright FIR emission could simply be a function of cloud compression.  As IV21 approached the disk, compression could have caused even a moderate amount of dust, already residing in the cloud, to heat and glow brightly.  This enhanced dust emission would suggest a much larger dust content than which actually resides in the cloud.  Regardless of the its origin, dust is required for the formation of molecules, including CO. 

Most of the molecular gas within the galaxy is located in the disk as high density regions commonly known as Giant Molecular Clouds (GMCs).  However, small amounts of CO have been detected in high-latitude clouds and the eight known IVMCs \citep{Weiss1999, MagnaniSmith2010}.  IV21 presents an nearby laboratory to study the process of molecular formation in a low-metallicity environment.  Is it possible that molecular hydrogen formed as a result of overdensity within HVCs and/or IVCs?  \citet{Federman1979} showed that for hydrogen gas in equilibrium, 10\% of the atomic hydrogen turns molecular if the \HI~ column density is larger than

\begin{equation}
\rm N_{HI,lim}=5.86\times10^{14}\left(\frac{\chi R_0}{2yGT^{1/2}n}\right)^m cm^{-2}
\end{equation}
where m$=1.4-1.6$ (theoretically determined), T is the temperature of the neutral hydrogen, n is the volume density of protons, R$_0$ is the photodissociation rate for the interstellar radiation field at $950 \Angstrom$, G is the standard \htwo~formation rate onto dust grains, $\chi$ is the scale factor for the intensity of the radiation field, and y is a scale factor for the \htwo ~formation rate dependent on metallicity and possibly other differences from the standard situation.  Similarly, in a search for CO within the densest parts of HVCs,  \citet{Wakker1997} estimated a range of HVC volume densities at which \htwo~would form.  These densities range from the largest of $\rm n=300\chi~ cm^{-3}$ for HVCs with atomic densities of $\rm \log\{N(HI)[\cm^{-2}]\}=20.0$ and $\rm m=1.6$ to the least dense of $\rm n=30\chi~ cm^{-3}$ for clouds with atomic densities of $\rm \log\{N(HI)[\cm^{-2}]\}=20.7$ with $\rm m=1.4$.  Overall, these implied volume densities are highly dependent on the chosen ISRF. 
  
Later, \citet{Richter2003} suggested that \htwo~formation must be taking place within the densest regions of HVCs/IVCs. A formation scenario requiring high density  is  required as \htwo~line self shielding is most likely not efficient in diffuse clouds with low molecular hydrogen fractions \citep{DraineBertoldi1996}. Using the relationship between the atomic and molecular volume densities from the \htwo~formation-dissociation equilibrium of \citet{Spitzer1978} and by assuming that the \htwo ~formation rate in the halo is the same as that within the disk, they found that an HVCs volume density was given by:

\begin{equation}
\rm  n_H\sim1.8\times10^6 \frac{N_{H2}}{N(HI)} \frac{\chi}{\phi} cm^{-3}.
\end{equation}
Here $\phi$ is the fraction of the HI gas along the line of sight which is related to the \htwo.  
Using this relation with our estimated \htwo~and \HI~ column densities for IV21 and assuming $\phi =1$, we derive a volume density of $\rm n_H\sim880\chi cm^{-3}$.  

To estimate the physical volume density of IV21, we need an estimate of the ISRF in the vicinity of the Sun and $\sim300\pc$ above the plane.  \citet{Wakker1997} created a model of the ISRF based on the \citet{Mathis1983} model, which assumed an emissivity from early type stars and the disk geometry described by \citet{Smith1978}.  \citet{Wakker1997} updated the disk geometry to a current Galactocentric radius of the Sun of $8.5~\kpc$.  At a height of $\sim 300~\pc$ above the plane their model predicts an ISRF of $\chi=0.25$.  Using this value, we estimate the volume density of IV21 to be $\rm n_H=220 ~\cm^{-3}$.  This density is higher than typically seen in Galactic IVCs containing \htwo, $\sim30~\cm^{-3}$.  However, following the relation from \citet{Federman1979}, which states that at least 10\% of the atomic hydrogen will become molecular for column densities larger than $\rm N_{HI,lim}$, then the molecular volume density of IV21 is at least $\rm n_{H_2}=0.05n_{H}$, or $\sim10~\cm^{-3}$.  This value is still lower than the CO $(1\rightarrow0)$ transition critical density of $\sim10^{3}~\cm^{-3}$.  Although \citet{Weiss1999} detected CO within IV21, it was primarily concentrated within the densest regions of the cloud. Given that PG1144+615 is located in a ``valley'' between two of IV21's densest regions, our estimated volume density is consistent with a density that is slightly lower in the direction of the background star. 

The cloud scale along the line of sight, $\rm s_c$, or ``depth'' can also be determined.  Given $\rm n_H s_c=$N(\HI), we find the depth of IV21 to be $\sim0.19$ pc. Although this scale estimate is lower than expected for a cloud that spans nearly 3 degrees, it also agrees with the ``valley'' location of PG1144+615.  Given that the distance to IV21 is $\sim300$~pc, each of these cores is on the order of $\sim5$~pc across.  Assuming the clumps are spherical, the maximum thickness of IV21 would also be on the order of $\sim5 \pc$.  Thus, our measurement reasonably describes a thin filamentary medium between the two dense cores. 

Due to the presence of molecular gas detected in IV21, we must ask if the cloud is currently harboring active star formation.  Figure \ref{grid1} presents the IRAS 100~$\mu$m and WISE 22~$\mu$m images of the IV21 region with the background star PG1144+615.  We find no evidence of star formation within the cloud from a visual inspection of the 22~$\mu$m WISE image.  A search of the WISE public source catalogs failed to find any embedded point sources within IV21.  Thus, we believe that IV21 has no ongoing star formation. 

\section{Conclusions}

Using FUSE and STIS absorption line spectra towards the sdB star PG1144+615, we have estimated the metallicity and \htwo~ content of the IVMC IV21. To do this, we combined FUV absorption line spectra with 21-cm line observations from GBT and EBHIS and infrared data sets from IRAS and WISE.  We conclude that: 

\noindent
(1) The atomic hydrogen component, as traced by the EBHIS data, and the bright FIR emission regions are well correlated throughout the IVMC. This suggests that IV21 has a substantial dust content.  There are still open questions regarding how such a large amount of dust became present within a sub-solar metallicity molecular cloud. 

\noindent
(2) Based on 23 interstellar absorption lines of eight ions combined with GBT 21-cm line observations, IV21 has a sub-solar metallicity of $\rm \log  (Z/Z_{\odot})=-0.43\pm0.12$~dex. 
Such a low metallicity is unusual for an IVC, especially for a cloud with such bright FIR emission.  We believe that IV21 is of a ``HVC impact origin'' in which a low metallicity cloud accreted in from extra-galactic/circum-galactic regions. Most likely, as the cloud progressed towards the lower Galactic halo it mixed with the local ISM, enriching the cloud to a metallicity consistent with those measured in HVCs. 

\noindent
(3) In agreement with IV21's sub-solar metallicity, its measured $I_{100 \mu m}/$N(\HI) ratio is only one-third of the usual solar neighborhood ratio of $\sim 1\times10^{-20} \rm ~(MJy~sr^{-1}~\cm^{2})$. Thus, IV21 the only cloud in the solar neighborhood known to have a significantly sub-solar metallicity. 

\noindent
(4) The properties of the molecular gas within IV21 were studied using the various \htwo~lines detected in the FUSE/FUV absorption line data.  An analysis of the first three J-levels concluded a molecular hydrogen column density of the cloud in the direction of PG1144+615 to be $\rm \log \{N(H_2)[cm^{-2}]\}=16.80^{+017}_{-0.26}$, lower than the values normally measured in IVMCs. Combining this result with the N(\HI) estimated by the GBT data, we estimate a hydrogen volume density of n$\rm _H=220~cm^{-3}$ and a depth of $\sim0.19$~pc.  Although the estimated atomic and molecular column densities differ from those normally seen in IVMCs ($\rm n_H\sim30~cm^{-3}$ and N(\htwo) $\sim10^{20}~\rm \rm cm^{-2}$) and the cloud's depth is lower than expected for a cloud which spans nearly 3 degrees, our results are reasonable given the ``valley'' location of PG1144+615 between two of IV21's densest clumps. 

Overall, IV21 presents an nearby laboratory which is ideal for studying the process of molecular formation in a low-metallicity environment.  In the future, we aim to combine this study with other metallicity studies of IVMCs to understand if they represent a special class of high-latitude clouds representing the mixing of HVCs and the lower Galactic halo.  Furthermore, the fact that CO is found within the cloud's densest cores \citep[e.g.][]{Weiss1999,MagnaniSmith2010}, implies that downward falling clouds may have an impact on Galactic star formation.  Regardless of where HVCs, IVCs, and IVMCs originate, their downward falling kinematics allow us to search for signatures of gas compression as they approach the disk and to explore the possibility that HVC impacts can influence star formation.

\noindent B.P.W and A.K.H acknowledge support from HST grant GO-12275 and NASA-ADP grant NNX11AD18G.  D.F acknowledges support from the grant NASA-AST-1108913 to B.P.W.  R.A.B and B.P.W acknowledge support for FUSE science from the NASA grants NAG5-12196 and ATP-NNX10AI70G.  Our work is based on data obtained by the NASA-CNES-CSA \textit{FUSE} mission operated by the Johns Hopkins University. Our results are also based on data from the EBHIS; based on observations with the 100-m telescope of the MPIfR (Max-Planck-Institut f\"{u}r Radioastronomie) at Effelsberg.  We thank the anonymous referee for their insightful comments. \\
\textit{FACILITIES:} \facility{HST (STIS)}, \facility{FUSE}, \facility{GBT}, \facility{Effelsberg}, and \facility{IRAS}.

\bibliography{hvc}
\end{document}